\documentstyle[prl,aps,times,amssymb,amsbsy,twocolumn]{revtex}  
\input{epsf.sty} 

\begin{document}    
\title{The temperature dependent behaviour of surface states in 
ferromagnetic semiconductors} 
\author{R.~Schiller\cite{email}} 
\address{Department of Physics and Astronomy, Northwestern University, Evanston,
         IL, 60208-3112, USA}
\author{W.~M\"uller and W.~Nolting} 
\address{Humboldt-Universit\"at zu Berlin, Institut f\"ur Physik,
         Invalidenstra{\ss}e 110, D-10115 Berlin, Germany}
\date{\today} 
\maketitle

\begin{abstract}
We present a model calculation for the temperature dependent behaviour of
surface states on a ferromagnetic local-moment film. The film is described
within the {\em s-f} model featuring local magnetic moments being exchange
coupled to the itinerant conduction electrons. The surface states are generated
by modifying the hopping in the vicinity of the surface of the film. In the
calculation for the temperature dependent behaviour of the surface states we are
able to reproduce both Stoner-like and spin-mixing behaviour in agreement with
recent (inverse) photoemission data on the temperature dependent behaviour of
a Gd(0001) surface state.
\end{abstract}

In the recent past many theoretical and experimental research works have been
focussed on the intriguing properties of rare earth metals and their compounds.
On the exeperimental side, this interest was aroused after Weller 
{\em et al.}~reported on the existence of the magnetically ordered Gd(0001)
surface at temperatures where the bulk Gadolinium is paramagnetic
\cite{WAG+85}. 
Since then, a variety of different experimental techniques has been applied to
the problem by different groups yielding values of the surface Curie temperature
enhancement, $\Delta T_C = T_C({\rm surface}) - T_C({\rm bulk})$, between 17K
and 60K \cite{RE86,RR87,TWW+93}. Contrary to the groups cited above Donath 
{\em et al.}~ using spin-resolved photoemission did not find any indication for
an enhanced Curie temperature at the Gd(0001) surface \cite{DGP96}, fueling the
controversial discussion. 

Concerning the interplay between electronic structure and 
exceptional magnetic properties at the Gadolinium surface 
a Gd(0001) surface state \cite{WF91,LHDH+91} is believed to play a crucial
role and its temperature dependent behaviour has been discussed intensely
\cite{DGP96,DDN98,FSK94,WSLM+96,DG98,BGPH+98,WD98}. 
Recently, the investigation of the correlation
between strain induced alteration of the surface electronic structure and
enhanced magnetization in Gd films has been addressed by a number of
experimental works \cite{WD98,WMIMA+98,WMAW+98,KWD99}. 
A thorough account on the
surface magnetism and the surface electronic structure of the
lanthanides has been given by Dowben {\em et al.} \cite{DMIL97}. 

Rare-earth materials are so-called local-moment systems, i.e.~the magnetic moment
stems from the partially filled {\em 4f}-shell of the rare-earth atom
being strictly localized at the ion site. Thus
the magnetic properties of these materials are determined by the
localized magnetic moments. On the other hand, the electronic properties like
electrical conductivity are borne by itinerant electrons in rather broad
conduction bands, e.g.~{\em 6s}, {\em 5d} for Gd. Many of the characteristics of
local-moment systems can be attributed to a correlation between the localized
moments and the itinerant conduction electrons. For this situation the {\em s-f}
model has been proven to be an adequate description. In this model, the
correlation  between localized moments and conduction electrons is represented
by an intraatomic exchange interaction.

In what follows we consider a film consisting of $n$ equivalent 
parallel layers. The lattice sites within the film are indicated by a greek
letter $\alpha$, $\beta$, $\gamma$, $\ldots$, denoting the layer, and by a latin
letter $i$, $j$, $k$, $\ldots$, numbering the sites within a given layer.
Each layer posseses two-dimensional translational symmetry, so for any site
dependent operator $A_{i\alpha}$ we have:
\begin{displaymath}
\left\langle A_{i\alpha} \right\rangle \equiv 
\left\langle A_\alpha \right\rangle .
\end{displaymath}
The Hamiltonian for the {\em s-f} model consists of three parts:
\begin{equation}
\label{ham}
{\cal H} = {\cal H}_s + {\cal H}_{f} + {\cal H}_{sf}.
\end{equation}
The first,
\begin{equation}
\label{h_s}
{\cal H}_s = \sum_{ij\alpha\beta\sigma} T^{\alpha\beta}_{ij} c^+_{i\alpha\sigma}
c_{j\beta\sigma},
\end{equation}
describes the itinerant conduction electrons as {\em s}-electrons with
$c^+_{i\alpha\sigma}$ ($c_{i\alpha\sigma}$) being the creation (annihilation)
operator of an electron with the spin $\sigma$ at the lattice site 
${\bf R}_{i\alpha}$. $T^{\alpha\beta}_{ij}$ are the hopping integrals.

The second part of the Hamiltonian represents the system of the localized 
{\em f}-moments and consists itself of two parts,
\begin{equation}
\label{h_f}
{\cal H}_f = -\sum_{ij\alpha\beta} J^{\alpha\beta}_{ij} {\bf S}_{i\alpha}
{\bf S}_{j\beta} -D_0\sum_{i\alpha}\left(S^z_{i\alpha}\right)^2,
\end{equation}
where the first is the well-known Heisenberg interaction. Here the 
${\bf S}_{i\alpha}$ are the spin operators of the localized magnetic moments,
which are coupled by the exchange integrals, $J^{\alpha\beta}_{ij}$.
The second contribution is a single-ion anisotropy term which arrises from the
necessity of having a collective magnetic order at finite temperatures, 
$T>0$ \cite{SN99a}. This anisotropy has been assumed to be uniform within the
film. The according anisotropy constant $D_0$ is typically smaller by some
orders of magnitude than the Heisenberg exchange integrals, 
$D_0\ll J^{\alpha\beta}_{ij}$.

In addition to the contribution of the {\em s}-electron system and the
contribution of the localized {\em f}-moments we have a third term which
accounts for an intraatomic interaction between the conduction electrons and the
localized {\em f}-spins:
\begin{equation}
\label{h_sf}
{\cal H}_{sf} = - \frac{J}{\hslash} \sum_{i\alpha} {\em S}_{i\alpha} 
\sigma_{i\alpha},
\end{equation}
where $J$ is the {\em s-f} exchange interaction and $\sigma_{i\alpha}$ is the
Pauli spin operator of the conduction electrons. In the case where $J<0$ the
Hamiltonian (\ref{ham}) is that of the so called {\em Kondo lattice}. However,
here we are interested in the case of positive {\em s-f} coupling ($J>0$) which
applies to the materials we are interested in.
Using the abbreviations,
\begin{displaymath}
S^{\sigma}_{i\alpha} = S^x_{i\alpha} + i z_{\sigma} S^y_{i\alpha}; \quad
z_{\uparrow(\downarrow)} = \pm 1,
\end{displaymath}
the {\em s-f} Hamiltonian can be written in the form:
\begin{equation}
\label{h_sf2}
{\cal H}_{sf} = - \frac{J}{2} \sum_{i\alpha\sigma} \left( z_{\sigma}
S^z_{i\alpha} n_{i\alpha\sigma} + S^{\sigma}_{i\alpha} c^+_{i\alpha -\sigma}
c_{i\alpha\sigma} \right).
\end{equation}

The problem posed by the Hamiltonian (\ref{ham}) can be solved by
considering the retarded single-electron Green function
\begin{eqnarray}
\label{green}
G^{\alpha\beta}_{ij\sigma} (E) & = & \left\langle\!\left\langle
c_{i\alpha\sigma} ; c^+_{j\beta\sigma} \right\rangle\!\right\rangle_E
\nonumber \\
& = & - {\rm i} \int\limits_0^{\infty} {\rm d}t\;{\rm e}^{-\frac{\rm i}{\hslash}
Et} \left\langle \left[ c_{i\alpha\sigma}(t),c^+_{j\beta\sigma}(0) \right]_+ 
\right\rangle,
\end{eqnarray}
which is related to the spectral density $S^{\alpha\beta}_{{\bf k}\sigma}(E)$
and the local density of states $\rho_{\alpha\sigma}(E)$ via the relations:
\begin{eqnarray}
\label{ftgreen}
G^{\alpha\beta}_{{\bf k}\sigma}(E) & = & \frac{1}{N} \sum_{ij} 
{\rm e}^{{\rm i}{\bf k}({\bf R}_i-{\bf R}_j)} G^{\alpha\beta}_{ij\sigma}(E),\\
\label{spectral}
S^{\alpha\beta}_{{\bf k}\sigma}(E) & = & - \frac{1}{\pi} {\rm Im} \,
G^{\alpha\beta}_{{\bf k}\sigma}(E+{\rm i}0^+), \\
\label{ldos}
\rho_{\alpha\sigma}(E) & = & \frac{1}{\hslash N} \sum_{\bf k} 
S^{\alpha\alpha}_{{\bf k}\sigma} (E).
\end{eqnarray}
Due to the translational symmetry of the films, the Fourier transformation 
(\ref{ftgreen}) has to be performed within the layers of the film. Accordingly, 
$N$ is the number of sites per layer, ${\bf k}$ is an in-plane wavevector from
the first two-dimensional Brillouin zone, and ${\bf R}_i$ represents the in-plane part of the
position vector, \mbox{${\bf R}_{i\alpha} = {\bf R}_i + {\bf r}_{\alpha}$}. 

The many-body problem that arises with the Hamiltonian (\ref{ham}) is far from
being trivial and a full solution is lacking even for the case of the bulk.
In a previous paper we have presented an approximate treatment of the special
case of a single electron in an otherwise empty conduction band \cite{SN99b}.
This solution, which holds for arbitrary temperatures, is based on the special
case of an empty conduction band interacting with a ferromagnetically 
saturated local-moment system ($T=0$), which can be solved exactly for both, 
the bulk case and the film geometry \cite{AE82,ND85,NMJR96,SMN97}. This exactly 
soluble limiting case gives the approximate solution for finite temperatures 
a certain trustworthiness. 

In the following formulas, we briefly recall the results of the 
calculations presented in \cite{SN99b,SMN97}. Due to the empty conduction 
band, we are considering throughout the whole paper, the Hamiltonian 
(\ref{ham}) can be split into an electronic part, ${\cal H}_s + {\cal H}_{sf}$, 
and a magnetic part, ${\cal H}_f$, which can be solved separately \cite{SN99b}. 
For the electronic part, 
\begin{equation}
\label{elpart}
{\cal H}^{el}= {\cal H}_s + {\cal H}_{sf},
\end{equation}
we employ the single-electron Green function (\ref{green}). The equation of
motion for this Green function $G^{\alpha\beta}_{ij\sigma}(E)$
can be formally solved 
by introducing the self-energy $M^{\alpha\beta}_{ij\sigma}(E)$, 
\begin{equation}
\label{selfen}
\left\langle\!\left\langle [ c_{i\alpha\sigma}, {\cal H}_{sf}]_- ;
c^+_{j\beta\sigma} \right\rangle\!\right\rangle_E = 
\sum_{m\mu} M^{\alpha\mu}_{im\sigma}(E) G^{\mu\beta}_{mj\sigma}(E),
\end{equation}
which contains all the information about correlation between the conduction band
and the localized moments. With the help of (\ref{selfen}) the equation of
motion for the single-electron Green function simply becomes, after
two-dimensional Fourier transform,
\begin{equation}
\label{formsol}
{\bf G}_{{\bf k}\sigma}(E) = \hslash \left( E \, {\bf I}-{\bf T}_{\bf k}-
{\bf M}_{{\bf k}\sigma}(E) \right)^{-1}.
\end{equation}
Here, ${\bf I}$ represents the $(n\times n)$ identity matrix and the matrices
${\bf G}_{{\bf k}\sigma}(E)$, ${\bf T}_{\bf k}$, and 
${\bf M}_{{\bf k}\sigma}(E)$ have as elements the layer-dependent functions
$G^{\alpha\beta}_{{\bf k}\sigma}(E)$, $T^{\alpha\beta}_{\bf k}$, and 
$M^{\alpha\beta}_{{\bf k}\sigma}(E)$, respectively. The necessary computation of
the self-energy $M^{\alpha\beta}_{{\bf k}\sigma}(E)$ involves the evaluation of
higher Green functions originating from the equation of motion of the original
single-electron Green function. For the sake of brevity we here omit the details
of the calculations. If the self-energy is assumed to be a local entity,
\begin{equation}
\label{localself}
M^{\alpha\beta}_{{\bf k}\sigma} (E) = -\frac{J}{2} \delta_{\alpha\beta}
m^{\alpha}_{\sigma}(E),
\end{equation}
it can be shown to have the structure:
\begin{equation}
\label{selfsol}
m^{\alpha}_{\sigma}(E)= \frac{Z^{\alpha}_{\sigma}(E)}{N^{\alpha}_{\sigma}(E)},
\end{equation}
where the numerator and the denominator, respectively, 
have the structure:
\begin{mathletters}
  \label{num_den}
  \begin{eqnarray}
    \label{num_den_z}
    Z^{\alpha}_{\sigma} & = & z_{\sigma} \hslash^2 \langle S^z_{\alpha}
    \rangle + \frac{J}{2} f^Z_1(\ldots) + \frac{J^2}{4} f^Z_2(\ldots) \\
    \label{num_den_n}
    N^{\alpha}_{\sigma} & = & \hslash^2 + \frac{J}{2} f^N_1(\ldots) + 
    \frac{J^2}{4} f^N_2(\ldots)
  \end{eqnarray}
\end{mathletters}
where the four functions $f^{Z,N}_{1,2}(\ldots)$ themselves depend
\footnote{For the explicit form of Eqs.~(\ref{num_den}) see \cite{SN99b}.} 
on the self-energy,
$m^{\alpha}_{\pm \sigma} (E)$, and on the layer-dependent coeffcients:
\begin{eqnarray}
  \label{coeff} 
  \kappa_{\alpha\sigma} & = &
  \langle S^{-\sigma}_{\alpha} S^{\sigma}_{\alpha} \rangle
  - \lambda^{(2)}_{\alpha\sigma} \langle S^z_{\alpha} \rangle,
  \nonumber \\
  \lambda^{(1)}_{\alpha\sigma} & = &
    \displaystyle \frac{\langle S^{-\sigma}_{\alpha}
    S^{\sigma}_{\alpha} S^z_{\alpha} \rangle + z_{\sigma}
    \langle S^{-\sigma}_{\alpha} S^{\sigma}_{\alpha} \rangle}
    {\langle S^{-\sigma}_{\alpha} S^{\sigma}_{\alpha} \rangle},
  \\
  \lambda^{(2)}_{\alpha\sigma} & = &
    \displaystyle \frac{\langle S^{-\sigma}_{\alpha}
    S^{\sigma}_{\alpha} S^z_{\alpha} \rangle - \langle S^z_{\alpha}
    \rangle \langle S^{-\sigma}_{\alpha} S^{\sigma}_{\alpha} \rangle}
    {\langle (S^z_{\alpha})^2 \rangle - \langle S^z_{\alpha} \rangle^2}.
    \nonumber
\end{eqnarray}
Taking into account Eqs.~(\ref{formsol})-(\ref{coeff}), we have now a closed
system of equations, provided that the {\em f}-spin correlation functions
appearing in Eqs.~(\ref{coeff}) are known. 

These can be evaluted by considering the magnetic subsystem, according to the
Hamiltonian ${\cal H}_f$ (Eq.~(\ref{h_f})). For all what concerns us in
this paper it is only interesting that, employing an RPA-type decoupling, 
a solution of the magnetic subsystem can be found\cite{SN99a}
and that it gives us all the necessary layer-dependent {\em f}-spin correlation
functions as a function of temperature from $T=0$ to the Curie temperature,
$T=T_C$ \cite{SN99b}. Mediated by Eqs.~(\ref{selfsol})-(\ref{coeff}), 
the {\em f}-spin correlation functions contain the whole temperature 
dependence of the electronic subsystem.

\begin{figure}[htbp]
  \centerline{\epsfxsize=\linewidth \epsffile{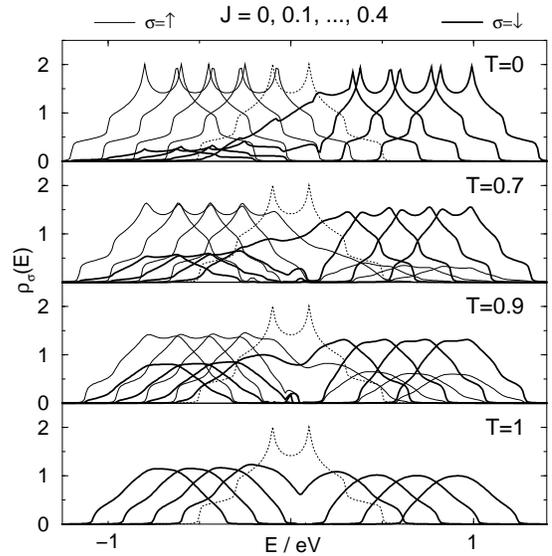}}
  \caption{Density of states $\rho_{\sigma}(E)$ of a s.c.-(100) double layer for
  different {\em s-f} interactions $J$ and different temperatures $T$ (in units
  of $T_C$). The dotted lines correspond to the case of vanishing {\em s-f}
  interaction, $J=0$, where there is no distinction between spin-$\uparrow$ and
  spin-$\downarrow$ electron, $\rho_{\uparrow}(E)=\rho_{\downarrow}(E)$.
  The spectra for the increasing values of the {\em s-f} interaction,
  $J=0.1,\,0.2,\,0.3,\,0.4$, can be associated
  by the increasing distances of their peaks from the $J=0$-spectrum.}
  \label{fig:1}
\end{figure}

To briefly recall the main results presented in our previous paper \cite{SN99b},
Fig.~\ref{fig:1} shows the density of states of a s.c.-(100) double layer for
different {\em s-f} interactions and different temperatures. In the case of
ferromagnetic saturation, $T=0$, we see that for the spin-$\uparrow$ electron
the density of states of the free case $J=0$ (dotted line) is just rigidly
shifted when the interaction is switched on. This is due to the impossibility of
the spin-$\uparrow$ electron to exchange its spin with the perfectly alligned
local-moment system. For the spin-$\downarrow$ electron in the case of
small {\em s-f} exchange coupling, $J>0$, a slight deformation of the free
density of states sets in. For intermediate and strong couplings
($J\gtrapprox0.2$), the density of states splits into two parts, 
corresponding to two different
spin exchange processes between the spin-$\downarrow$ electron and the localized 
{\em f}-spin system. The higher energetic part represents a polarization of the
immediate spin neighbourhood of the electron due to the repeated emission and
reabsorption of magnons. The corresponding polaron-like quasiparticle is called
the magnetic polaron. The low-energetic part of the spectrum is a scattering
band which can be explained by the simple emission of a magnon by the
spin-$\downarrow$ electron without reabsorption, but necessarily connected 
with a spin flip of the electron \cite{SMN97}.

For $T>0$ we see in Fig.~\ref{fig:1} that with
increasing temperature for the spin-$\downarrow$ electron 
spectral weight is transferred from the high-energetic polaron peak to the
low-energetic scattering peak. On the other hand, for the spin-$\uparrow$
electron we have the effect that for finite temperatures an additional peak
rises at the high energetic side of the spectrum. This rise with increasing
temperature is fueled by the loss of spectral weight on the low-energetic peak.
The high-energetic peak simply represents the ability of the spin-$\uparrow$
electron to exchange its spin with a not perfectly aligned local-moment system,
$T>0$ \cite{SN99b}. As a result of the shifts of spectral weight occuring
for both, the spin-$\uparrow$ and spin-$\downarrow$ spectra, the spectra for the
two spin directions approach each other with increasing temperature. In the
limiting case of $T\rightarrow T_C$ the system has eventually lost its ability to
distinguish between the two possible spin directions because of the loss of
magnetization of the underlying local-moment system. Hence, for $T=T_C$ the
densities of states of the spin-$\downarrow$ and the spin-$\uparrow$ electron
are equal. Another feature which can be seen in Fig.~\ref{fig:1} is
that while the positions of the quasiparticle subbands stays pretty much the
same we observe a narrowing of the bands with increasing temperature.
This can especially well be seen for the case of temperature evolvement of
spectrum of the spin-$\downarrow$ electron for the case of $J=0.2$. 
Whereas the scattering band and the polaron band are, for the case of $T=0$, 
still merged, in the case of $T=T_C$ both bands are clearly separated.

\begin{figure}[htbp]
  \centerline{\epsfxsize=\linewidth \epsffile{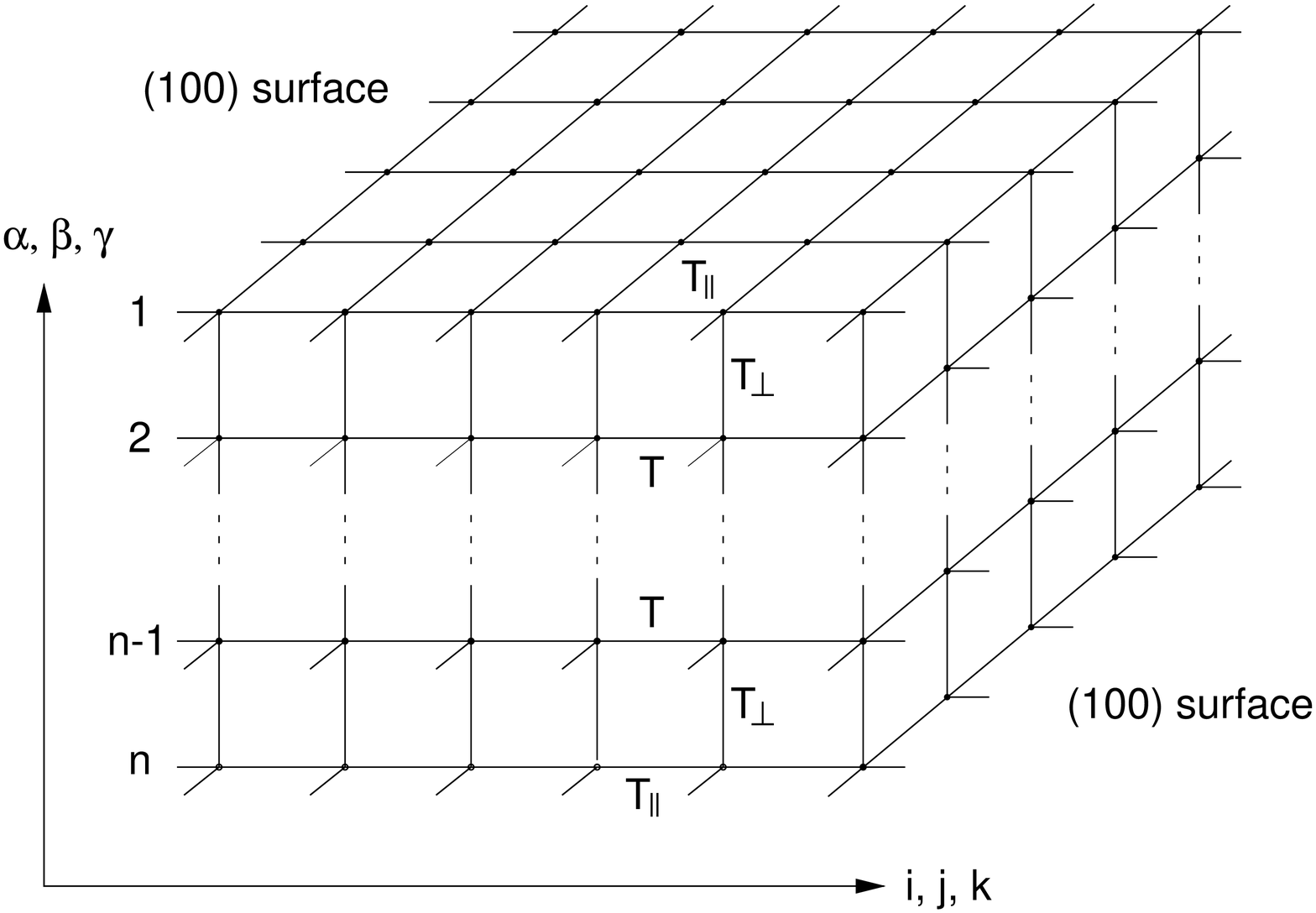}}
  \caption{Model of an $n$-layer film with s.c. structure. The nearest neighbour
  hopping is assumed to be $T_{\parallel}$ within the surface layer,
  $T_{\perp}$ between the surface layer and the layer nearest to the
  surface layer, and $T$ within and between inner layers.}
  \label{fig:2}
\end{figure}

In this paper we are interested in surface states and their temperature
dependent behaviour. Surface states occur
in the spectral density at energies different from the bulk energies 
and are localized in the vicinity of the surface of a crystal. 
The theory presented above 
is applied to a s.c. film consisting of $n$ layers oriented parallel to the
(100)-surface as drawn schematically in Fig.~\ref{fig:2}. 

The electron hopping in Eq.~(\ref{h_s}) is restricted to nearest neighbours,
\begin{equation}
\label{hopp_nn}
T^{\alpha\beta}_{ij} = \delta^{\alpha\beta}_{i,j\pm\Delta} T^{\alpha\alpha}
+\delta^{\alpha,\beta\pm 1}_{ij} T^{\alpha\beta},
\end{equation}
where $\Delta$ stands for the nearest neighbours within the same plane,
$\Delta=(0,1),\,(0,\bar{1}),\,(1,0),\,(\bar{1},0)$, and
$\delta^{\alpha\beta}_{ij}\equiv\delta_{\alpha\beta} \delta_{ij}$.
$T^{\alpha\beta}$ is the hopping between the adjacent layers $\alpha$ and 
$\beta$ and $T^{\alpha\alpha}$ is the hopping between nearest neighbours 
within the layer $\alpha$. In order to study surface states we vary the
electron hopping within the vicinity of the surface,
\begin{mathletters}
\label{hopping}
\begin{equation}
\label{hopping1}
T^{\alpha\beta} = \left( \begin{array}{cccccc} 
T_{\parallel} & T_{\perp} & 0 & & \!\!\!\!\!\cdots\,\,\,\,\, & 0 \\
T_{\perp} & T & T & & & \vdots \\ 0 & T & & \ddots & & \\
& & \ddots & & T & 0 \\ \vdots & & & T & T & T_{\perp} \\ 
0 & \,\,\,\,\,\cdots\!\!\!\!\!\ & & 0 & T_{\perp} & T_{\parallel}
\end{array} \right),
\end{equation}
according to Fig.~\ref{fig:2}, and with
\begin{equation}
\label{hopping2}
T_{\parallel}=\epsilon_{\parallel}T, \qquad
T_{\perp}=\epsilon_{\perp}T.
\end{equation}
\end{mathletters}
Here, $\epsilon_{\parallel}$ and $\epsilon_{\perp}$ are considered as model
parameters. In reality the variation of the hopping integrals in the vicinity of
the surface may be caused e.g.~by a relaxation of the interlayer distance.
According to the scaling law $T \sim r^{-5}$ for the {\em d}-electrons
\cite{Pap86} a relatively small top-layer relaxation $\Delta r / r$ 
may result in a strong change of the hopping integral $T$. 
Thus e.g.~a relaxation of the Gd(0001) surface layer of 3-6\%
(cf.~\cite{EAOT+95} and references therein) would yield a modification 
of the hopping integrals of of up to 30\%. 

In a previous paper we have been dealing with surface states for the exactly
solvable case of a single electron in an otherwise empty conduction band and a
ferromagnetically saturated {\em f}-spin system, $T=0$ \cite{SMN98}.
For this special case we have shown that modifying the hopping in the vicinity
of the surface according to Eqs.~(\ref{hopping}) leads to the appearance of
surface states in the local spectral density $S^{\alpha\alpha}_{{\bf k}\sigma}$.
Modifying the hopping within the surface layer by more than 25\%, i.e.
$\epsilon_{\parallel}\lessapprox\frac{3}{4}$ or 
$\epsilon_{\parallel}\gtrapprox\frac{5}{4}$,
while keeping all the other hopping integrals unchanged results in a single
surface state at the lower or the upper edge of the bulk band 
\cite{SMN98,MSN99}. This surface
state first emerges at the $\bar{\Gamma}$- and at the $\bar{\rm M}$-point from
the bulk band and from there spreads for larger modifications of
$\epsilon_{\parallel}$ to the rest of the Brillouin zone. 

On the other hand, when the hopping within the first layer remains constant,
$\epsilon_{\parallel}\equiv 1$, but the hopping between the first and the second
layer is significantly increased, $\epsilon_{\perp}\gtrapprox \sqrt{2}$, 
then two surface states split off one on each side of the bulk band. In this
case the emergence of the surface states from the bulk band is 
${\bf k}$-independent. Both types of surface states for the special
case of $T=0$ can be observed at the single bulk band and on the high-energetic 
polaron band for the case of the spin-$\uparrow$ electron and the
spin-$\downarrow$ electron, respectively. 

\begin{figure}[htbp]
  \centerline{\epsfxsize=\linewidth \epsffile{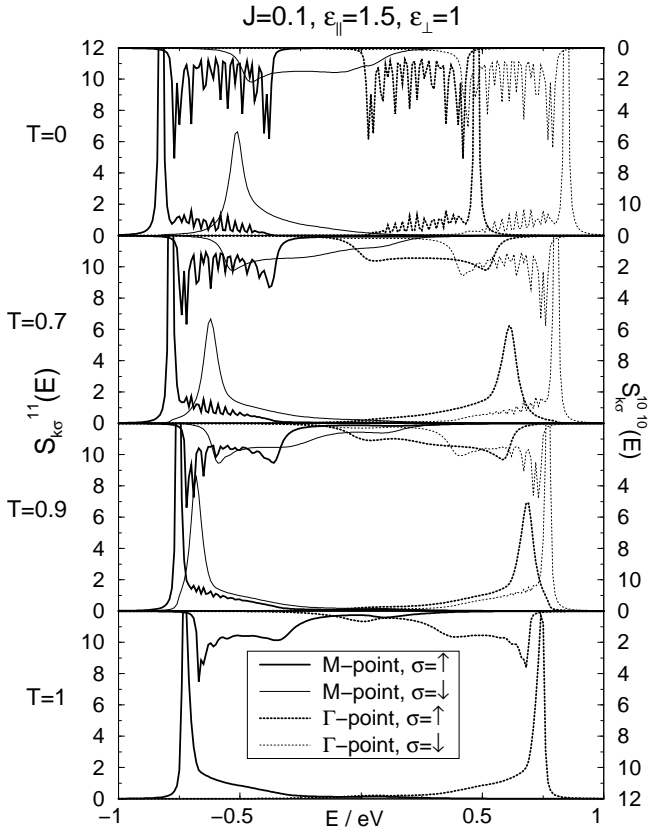}}
  \caption{Local density of states, $S_{{\bf k}\sigma}^{11}(E)$ of the first layer
  of a 20-layer s.c.-(100)-film at the $\bar{\Gamma}$-point and the 
  $\bar{\rm M}$-point for both spin directons and different temperatures,
  $T=0,\,0.7,\,0.9,\,1$ (in units of $T_C$) and modified hopping within the
  first layer, $\epsilon_{\parallel}=1.5$, while $\epsilon_{\perp}\equiv 1$.
  Upside down but in the same sacale, the local spectral density of a central 
  layer, $S_{{\bf k}\sigma}^{10\,10}(E)$, is displayed for the
  same ${\bf k}$-points and spin directions.}
  \label{fig:3}
\end{figure}

\begin{figure}[htbp]
  \centerline{\epsfxsize=\linewidth \epsffile{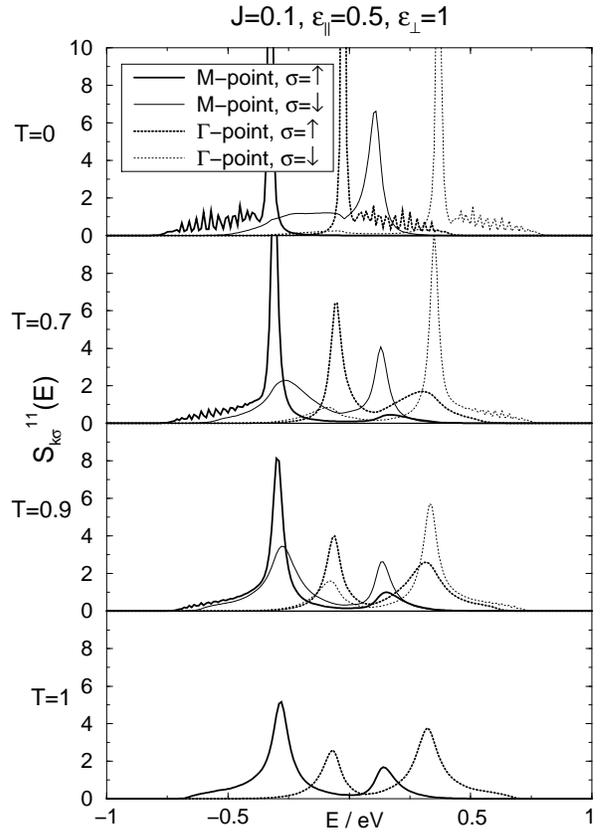}}
  \caption{Local density of states, $S_{{\bf k}\sigma}^{11}(E)$ of the first layer
  of a 20-layer s.c.-(100)-film at the $\bar{\Gamma}$-point and the 
  $\bar{\rm M}$-point for both spin directons and different temperatures,
  $T=0,\,0.7,\,0.9,\,1$ (in units of $T_C$) and modified hopping within the
  first layer, $\epsilon_{\parallel}=0.5$, while $\epsilon_{\perp}\equiv 1$.}
  \label{fig:4}
\end{figure}

\begin{figure}[htbp]
  \centerline{\epsfxsize=\linewidth \epsffile{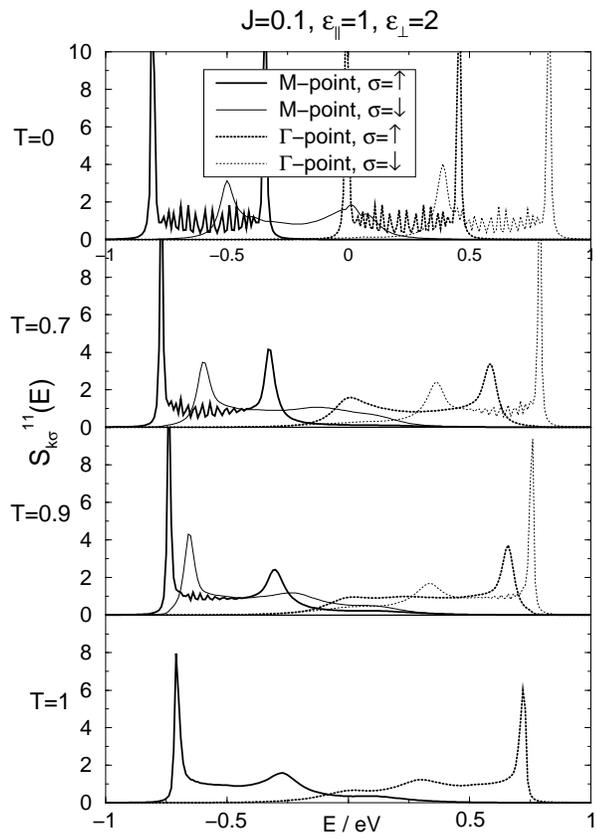}}
  \caption{Same as Fig.~\ref{fig:4} for modified hopping between the first and
  the second layer, $\epsilon_{\perp}=2$, while $\epsilon_{\parallel}\equiv 1$.}
  \label{fig:5}
\end{figure}

Here, we are interested in the possible variations of the $T=0$ surface states
when going to finite temperatures. Figs.~\ref{fig:3} to \ref{fig:5} show the
temperature dependence of surface states for the different possible variations
of the hopping in the vicinity of the surface. All the calculations for
Figs.~\ref{fig:3} to \ref{fig:5} have been performed for a 20-layer s.c. film
cut parallel to the (100)-plane of the s.c. crystal. For much thinner films 
than 20 layers the calculation of surface states becomes meaningless since
there is no real bulk-like environment in the center of the film to compare the
electronic states at the surface to. So it is desirable to calculate thicker 
films for the discussion of surface states. The choosen thickness of our model
film is basically a compromise between computational accuracy and computational 
time. The parameters for the
uniform hopping according to Eqs.~(\ref{hopping}) and for the {\em s-f} exchange
interaction are $T=-0.1$ and $J=0.1$, respectively.

For our calculations we have employed a modification of the 
hopping in the surface layer and between the surface layer and the adjacent
layer of 50\% ($\epsilon_{\parallel}=0.5,\,1.5$) and 100\%
($\epsilon_{\perp}=2$), respectively (cf.~Eq.~(\ref{hopping2})). Such a drastic
modification of the hopping integrals in the vicinity of the surface is rather
unlikely to occur in reality. However, as has been shown in \cite{SMN98} 
the actual peak position of a 
surface states depends only weakly on the variation of 
$\epsilon_{\parallel}$ and $\epsilon_{\perp}$, respectively. The selected strong
variation of the hopping parameters $T_{\parallel}$ and $T_{\perp}$
give rise to pronounced surface states which 
enable us to more clearly see the qualitative behaviour of the surface states 
as a function of temperature.

In Fig.~\ref{fig:3} we have the case that the hopping within the surface layer
is enhanced by 50\%, leading to the existence of a surface state on the outer
edge of the bulk dispersion. This surface state can most clearly be seen at the
$\bar{\Gamma}$-point and the $\bar{\rm M}$-point in the two-dimensional Brillouin
zone. In Fig.~\ref{fig:3}, the spectral density of the first layer, 
$S_{{\bf k}\sigma}^{11}(E)$, for both of these points is displayed 
as a function of energy. If the temperature is increased, we see that 
the position of the spin-$\uparrow$ and of the spin-$\downarrow$ surface 
states approach each other in a Stoner-like fashion until both peaks are 
equivalent for $T=T_C$. The oscillations in the spectral densities, which can
be seen in Fig.~\ref{fig:3} around -0.6eV, 0.3eV, and 0.7eV are due to the
finite thickness of our model film, as are the respective oscillations in
Figs.~\ref{fig:4} and \ref{fig:5}.

Also in Fig.~\ref{fig:3}, but upside down, the 
spectral density of one of the central layers of the  20-layer film, 
$S_{{\bf k}\sigma}^{10\,10}(E)$, can be seen again for the 
$\bar{\Gamma}$-point and the $\bar{\rm M}$-point and both spin directions
indicating that the positions of the surface states visible in 
$S_{{\bf k}\sigma}^{11}(E)$ lie outside of the bulk spectrum of the crystal
for all temperatures. The same is valid for the spectra displayed in
Figs.~\ref{fig:4} and \ref{fig:5}. In these figures, however, the local spectral
densities of the central layers have been omitted for clarity.

In Fig.~\ref{fig:4}, the temperature dependence of
surface states is documented for the case where the hopping within the first
layer is reduced by 50\%, $\epsilon_{\parallel}=0.5$ ($\epsilon_{\perp}\equiv1$).
In this case we observe a spin-mixing behaviour where the positions of
the spin-$\uparrow$ and of the spin-$\downarrow$ surface states stay the
same when the temperature is risen but spectral weight is being 
transferred between the different peaks. This results in equal populations
of the spin-$\downarrow$ and the spin-$\uparrow$ peaks at $T=T_C$.

To round up the picture we see in Fig.~\ref{fig:5} the case where the hopping
between the first and the second layer is modified, $\epsilon_{\perp}=2$,
while the hopping within the first layer remains equal to the uniform hopping
within the film, $\epsilon_{\parallel}\equiv 1$. Here we have for $T=0$ two
surface states, one on each side of the bulk spectrum. When the temperature is
switched on the surface states on the outer side of the bulk dispersion behave
Stoner-like, while the surface states on the inner side of the bulk dispersion 
exhibit a spin-mixing behaviour. 

Apparently, our model is able to reproduce a Stoner like collapse of the
spin-$\downarrow$ and the spin-$\uparrow$ peak positions for $T_C$ as well as a
spin-mixing behaviour. In the spin-mixing case there are two peaks which
both have a majority- and a minority-spin contribution. When the temperature is
increased, the spectral weight of these contributions is altered until for
$T=T_C$ for each peak the spin-$\downarrow$ and the spin-$\uparrow$
contributions have the same spectral weights.

In being able to reproduce both Stoner-like and spin-mixing behaviour, 
depending on the variation of the hopping and the position in the Brillouin
zone, our model calculations are in harmony
with more recent (inverse) photoemission studies on Gd(0001) films
which abandoned the grasp of earlier works that the temperature dependent 
behaviour of the Gd(0001) surface state has to be either Stoner-like or of
spin-mixing type \cite{DG98,BGPH+98}. Especially, it has been shown here 
that for certain parameters it is possible to observe both kinds of 
behaviour at the same time (see Fig.~\ref{fig:5}). This feature of our model
calculation seems to be in strong agreement with a scenario proposed by 
Donath and Gubanka \cite{DG98}.

\acknowledgments 
This work was supported by the Deutsche Forschungsgemeinschaft 
within the Sonderforschungsbereich 290 (``Metallische d\"unne Filme: 
Struktur, Magnetismus, und elektronische Eigenschaften''). 
One of the authors (R.S.) gratefully acknowledges the support by the German 
National Merit Foundation (``Studienstiftung des deutschen Volkes'').


\end{document}